\documentstyle[11pt,newpasp,epsf,twoside]{article}
\newcommand{\wavu}{ $ {\rm m ^ {-1} } $ }

\newcommand{\dmu}{${\rm pc ~ cm ^ {-3}}$\,}

\newcommand{\avbetameas}{\mbox{$ {\rm \langle \beta _{meas} \rangle } $}}
\newcommand{\nd}{\mbox{$ \nu _d $\,}}
\newcommand{\td}{\mbox{$ \tau _d $\,}}

\begin{document}

\title{Constraints on Interstellar Plasma Turbulence Spectrum from Pulsar Observations at the Ooty Radio Telescope} 

\author{N. D. Ramesh Bhat}
\affil{Max-Planck-Institut f\"ur Radioastronomie, Bonn, Germany}

\author{Yashwant Gupta, A. Pramesh Rao}
\affil{National Centre for Radio Astrophysics, Pune, India}

\section{Introduction} 

Refractive Interstellar Scintillation (RISS) effects on pulsar signals are powerful techniques 
for discriminating between different models that have been proposed for the power spectrum of 
plasma density fluctuations in the Interstellar Medium (ISM; e.g. Rickett 1990).
The nature of the spectrum is considered to be a major input for understanding the underlying 
mechanism of interstellar plasma turbulence.
Data from our long-term pulsar scintillation observations using the Ooty Radio Telescope (ORT) 
at 327 MHz are used to investigate the nature of the spectrum in the Local Interstellar Medium 
(LISM; region within $\sim$ 1 kpc of the Sun). Dynamic scintillation spectra were obtained for 
18 pulsars in the DM range 3$-$35 \dmu at $\sim$10$-$100 epochs spanning $\sim$100$-$1000 days
during 1993$-$1995 (Bhat et al. 1999).
From these observations, various scintillation properties and the ISM parameters are estimated 
with accuracies much better than that which has been possible from most earlier data.
The time series of parameters, $viz.$, decorrelation bandwidth (\nd), scintillation time scale 
(\td) and the drift slope of intensity scintillation patterns, and pulsar flux density 
are used to study various observable effects of Interstellar Scintillation, based on which the 
spectral form is inferred over the spatial scale range $\sim 10^7$ m to $\sim10^{13}$ m.

\section{Results and Conclusions} 

The main results from the present study are:

\begin{enumerate}
\item
Observations show large-amplitude modulations of the {\it diffractive scintillation} observables 
(\nd and \td) and flux.
The measured depths of modulations are considerably larger than the predictions of models 
based on a thin-screen scattering geometry and a simple Kolmogorov form of density spectrum.
\item
The statistical properties of diffractive and refractive scattering angles are used to 
obtain precise estimates of the spectral slope ($\beta$) over the spatial scale range $\sim 
10^7$ m to $\sim 10^{11}$ m. 
While for 12 pulsars, $\beta$ is found to be consistent with the Kolmogorov index 
(at $ \pm $ 2-$\sigma$ levels), it is found to be significantly larger for 6 pulsars ($ 11/3 < \beta < 4 $). \\
\item
From the anomalous scintillation behaviour of {\it persistent drift slopes} lasting over many 
months, we infer the power levels at spatial scales $\sim 10^{12} - 10^{13}$ m ($i.e.,$ 1$-$2 
orders of magnitude larger than refractive scales) to be 2$-$3 orders of magnitude larger than 
that expected from a Kolmogorov form of spectrum (Figure 1).
\end{enumerate}

\begin{figure}
\plotfiddle{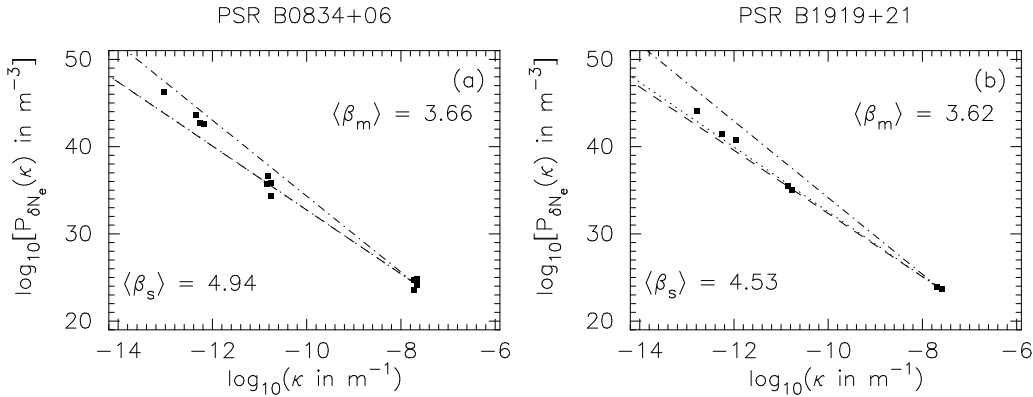}{2.15in}{270}{65}{65}{-265}{265}
\caption[]{Electron density power spectra for pulsars PSR B0834+06 and PSR B1919+21, 
determined using various observable effects of interstellar scattering seen in Ooty 
pulsar observations. The dashed line corresponds to the average slope (\avbetameas) 
determined by power levels at diffractive and refractive scales. 
The dotted line represents the Kolmogorov scaling ($\beta = 11/3$) and the 
dot-dashed line $\beta = 4$.}
\end{figure}

Although, there are various methods of investigating the nature of the spectrum, measurements based 
on a particular method give similar implications for the spectral form.  
A careful consideration of all the available results from the literature and our observations leads 
to the picture of a {\it Kolmogorov-like} spectrum ($\beta \approx 11/3$) in the spatial scale range 
$\sim 10^6$ m to $\sim 10^{11}$ m, with a low wavenumber enhancement (at $ \kappa \sim 10^{-12} - 
10^{-13}$ \wavu).
This is not quite in agreement with the results from some of the recent literature, where the spectrum 
is suggested to be an approximately power-law over the wide range $\sim 10^8$ m to $\sim 10^{13}$ m
(cf. Armstrong et al. 1995).
Further, for several nearby pulsars (distance $\sim$ 200$-$700 pc), the spectrum is found to be somewhat 
steeper ($11/3 < \beta < 4$), and there is a weak, systematic trend for a decrease in the spectral slope 
($\beta$) with distance.

A more complete description of this work can be found in Bhat N.D.R., Gupta Y., Rao A.P. 1999, ApJ, 514, 249.

\end{document}